\documentclass[prl, reprint, amsmath, amssymb, aps, preprintnumbers, longbibliography]{revtex4-2}
\usepackage{dcolumn}
\usepackage{bm}
\usepackage{kbordermatrix}
  
\usepackage[colorlinks=true,linkcolor=black, citecolor=black,
urlcolor=black]{hyperref}
\usepackage{xcolor}
\usepackage{amstext}
\usepackage{amssymb}
\usepackage{amsmath}
\usepackage{graphicx}
\usepackage{color}
\usepackage{caption}
\usepackage{textcomp}
\usepackage{tikz}
\usepackage{graphicx}
\usepackage{float}
\usepackage{subcaption}
\usepackage{xcolor}
\usepackage{nicematrix}

\usepackage{multirow}
\newcommand{\beq}{\begin{equation}}
\newcommand{\eeq}{\end{equation}}
\newcommand\beqa{\begin{eqnarray}}
\newcommand\eeqa{\end{eqnarray}}

\renewcommand{\le}{\leqslant}
\renewcommand{\leq}{\leqslant}
\renewcommand{\ge}{\geqslant}
\renewcommand{\geq}{\geqslant}

\newcommand{\wone}{\vcenter{\hbox{$\mathord{\includegraphics[width=4.5ex]{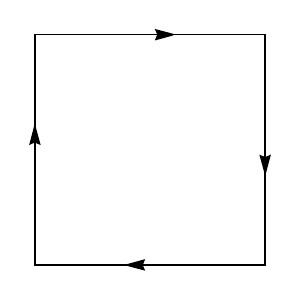}}$}}}
\newcommand{\wtwo}{\vcenter{\hbox{$\mathord{\includegraphics[width=6.5ex]{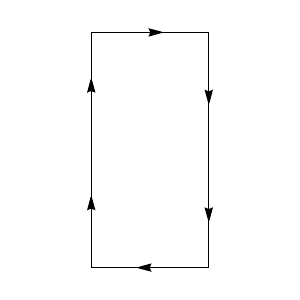}}$}}}
\newcommand{\wthree}{\vcenter{\hbox{$\mathord{\includegraphics[width=4.5ex]{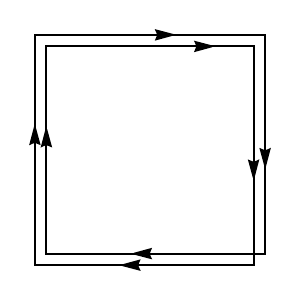}}$}}}
\newcommand{\wfour}{\vcenter{\hbox{$\mathord{\includegraphics[width=6.5ex]{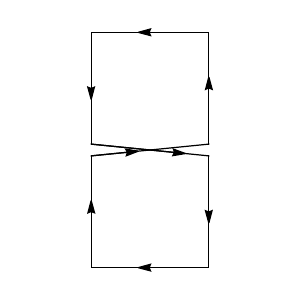}}$}}}
\newcommand{\wfive}{\vcenter{\hbox{$\mathord{\includegraphics[width=6.5ex]{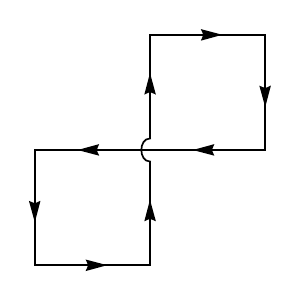}}$}}}
\newcommand{\wsix}{\vcenter{\hbox{$\mathord{\includegraphics[width=6.5ex]{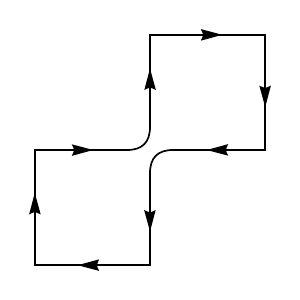}}$}}}
\newcommand{\lineone}{\vcenter{\hbox{$\mathord{\includegraphics[width=5.5ex]{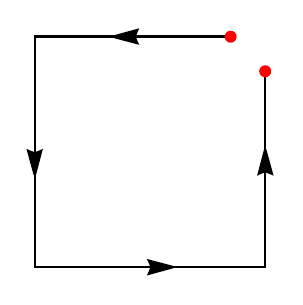}}$}}}
\newcommand{\linetwo}{\vcenter{\hbox{$\mathord{\includegraphics[width=5.5ex]{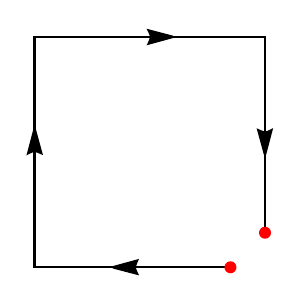}}$}}}
\newcommand{\linethree}{\vcenter{\hbox{$\mathord{\includegraphics[width=5.5ex]{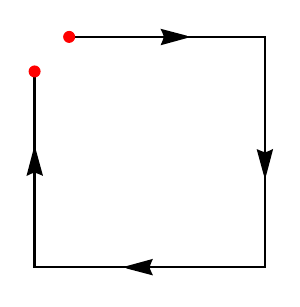}}$}}}
\newcommand{\linefour}{\vcenter{\hbox{$\mathord{\includegraphics[width=5.5ex]{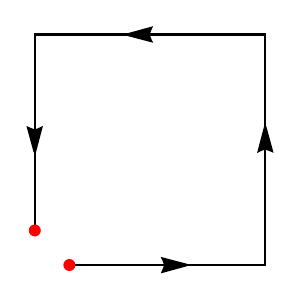}}$}}}
\newcommand{\linefive}{\vcenter{\hbox{$\mathord{\includegraphics[width=5.5ex]{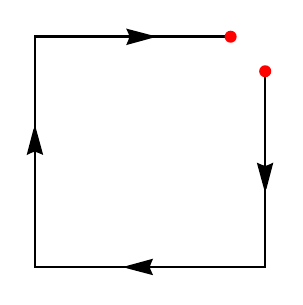}}$}}}
\newcommand{\linesix}{\vcenter{\hbox{$\mathord{\includegraphics[width=5.5ex]{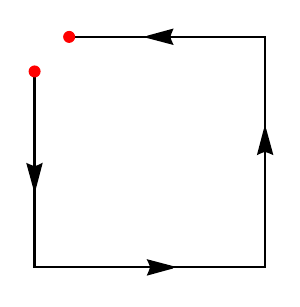}}$}}}
\newcommand{\lineseven}{\vcenter{\hbox{$\mathord{\includegraphics[width=5.5ex]{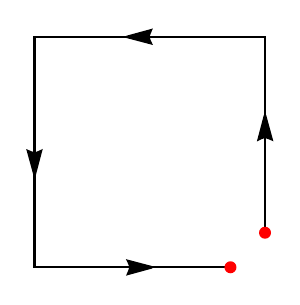}}$}}}
\newcommand{\lineeight}{\vcenter{\hbox{$\mathord{\includegraphics[width=5.5ex]{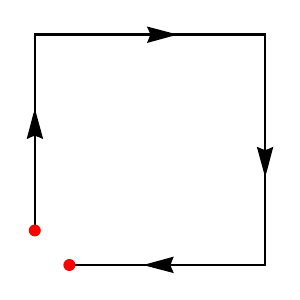}}$}}}

\newcommand{\backtrackposone}{\mathord{\includegraphics[width=6.5ex]{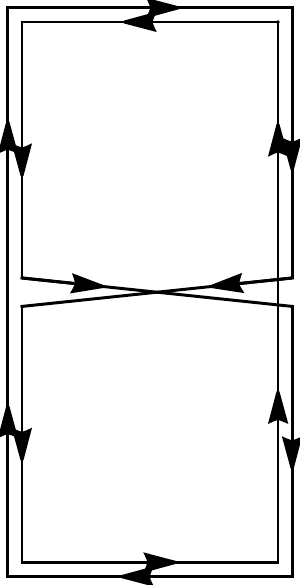}}}
\newcommand{\backtracknegone}{\mathord{\includegraphics[width=6.5ex]{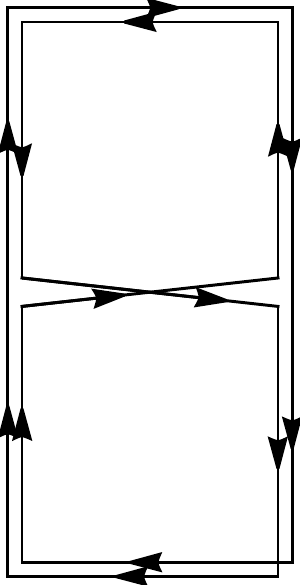}}}
\newcommand{\backtracknegtwo}{\mathord{\includegraphics[width=6.5ex]{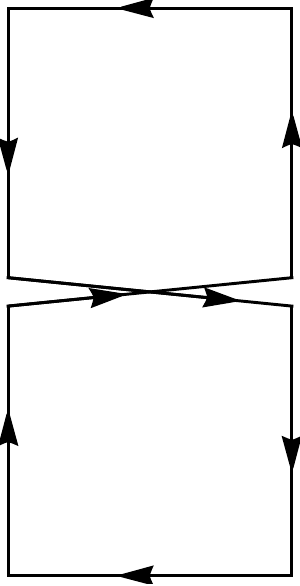}}}
\newcommand{\uplong}{\mathord{\includegraphics[width=6.5ex]{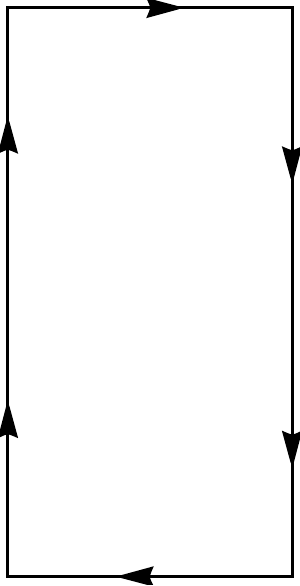}}}
\renewcommand{\kbldelim}{(}
\renewcommand{\kbrdelim}{)}
\newcommand{\pathone}{\mathord{\includegraphics[width=4.ex]{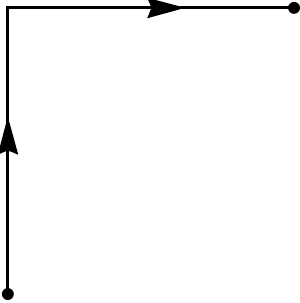}}}
\newcommand{\pathtwo}{\mathord{\includegraphics[width=4.ex]{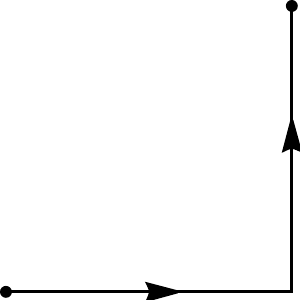}}}
\newcommand{\tr}{\text{tr}\,}

\begin{document}

\makeatletter
     \@ifundefined{usebibtex}{\newcommand{\ifbibtexelse}[2]{#2}} {\newcommand{\ifbibtexelse}[2]{#1}}
\makeatother


\title{Bootstrap for Lattice Yang-Mills theory}


\author{Vladimir Kazakov   {\it and}  Zechuan Zheng}
\email{kazakov $\bullet$ lpt.ens.fr}
\email{zechuan.zheng.phy $\bullet$ gmail.com}

\affiliation{ Laboratoire de physique de l'\'Ecole normale sup\'erieure, ENS, Universit\'{e} PSL, CNRS, Sorbonne Universit\'e,  Universit\'e Paris Cit\'e, 24 rue Lhomond, F-75005 Paris, France}

\affiliation{Interdisciplinary Scientific Center Poncelet (CNRS UMI 2615)} 


\begin{abstract}
We study the \(SU(\infty)\) lattice Yang-Mills theory at the dimensions \(D=2,3,4\)  via the numerical bootstrap method. It combines the LEs, with a cut-off \(L_{\mathrm{max}}\) on the maximal length of loops, and positivity conditions on certain matrices of Wilson loop averages. Our algorithm is inspired by the pioneering paper of P.Anderson and  M.Kruczenski but it is significantly more efficient, as it takes into account the symmetries of the lattice theory and uses the relaxation procedure in line with our previous work on matrix bootstrap.    We thus obtain rigorous upper and lower bounds on the plaquette average at various couplings and dimensions.      For \(D=4,\,\, L_{\mathrm{max}}=16\) the lower bound data appear to be close to the Monte Carlo data in the strong coupling phase and the upper bound data in the weak coupling phase reproduce well the 3-loop perturbation theory. Our results suggest that this bootstrap approach can provide a tangible alternative to the, so far uncontested,  Monte Carlo approach.            
\end{abstract} 

\maketitle

\section{Introduction}

The \(SU(N_c) \)  Yang-Mills (YM) lattice gauge theory (LGT) is a fundamental ingredient of modern particle physics. Its most illustrious applications are the Standard Model and, in particular, the Quantum Chromodynamics. Nowadays, most of the non-perturbative computations in Yang-Mills theory are done by Monte Carlo (MC) simulations for the lattice formulation of YM theory. Combined with the perturbation theory (PT) \cite{Bali:2014fea, Luscher:2014mka, DallaBrida:2017tru, DelDebbio:2018xpi, Perez:2017jyq} and RG tools,  MC methods have had a huge success, especially in the recent couple of decades, due to the development of supercomputers.  It allowed us to compute with a reasonable precision certain masses of hadrons and the S-matrix elements in QCD, reproducing the experimental data \cite{Gattringer:2010zz, Weisz:2012}.   However, the absence of any systematic non-perturbative ``analytic"  alternative to MC is, practically and intellectually, somewhat uncomfortable.  Moreover, the MC method has its inherent limitations: statistical errors, finite lattice size, high numerical cost of inclusion of dynamical quarks, difficulties in treating finite baryon density, and the real-time dynamics.

An interesting alternative for the study of the LGT is provided by Makeenko-Migdal loop equations (LE)~\cite{Makeenko:1979pb, Migdal:1983qrz} for Wilson loop averages (WA). An early attempt at numerical study of LE in the large \(N_c\), 't~Hooft limit was proposed in  \cite{Jevicki:1982jj, Jevicki:1983wu, Rodrigues:1985aq, Koch:2021yeb}, in the form of minimization of effective action in the loop space. A more recent brave attempt to bootstrap the LE, combining them with certain positivity conditions~\cite{Anderson:2016rcw, Anderson:2018xuq} revived hopes of a more analytic approach. Slightly later, a similar bootstrap method was proposed in \cite{Lin:2020mme} for the multi-matrix models. In our work
\cite{Kazakov:2021lel}
we significantly improved matrix bootstrap by introducing a ``relaxation" procedure and applied it to an ``analytically unsolvable" large \(N\) 2-matrix model, with remarkable efficiency and precision, noticeably exceeding those of MC for the same model~\cite{Jha:2021exo}.

These developments have been considerably inspired by the success of the bootstrap approach to  CFTs~~\cite{Rattazzi_2008, Poland:2018epd}  and \(S\)-matrices in massive QFTs~\cite{Paulos:2016fap, Guerrieri:2018uew, EliasMiro:2021nul}.

\begin{figure}
  \includegraphics[width=.5\textwidth]{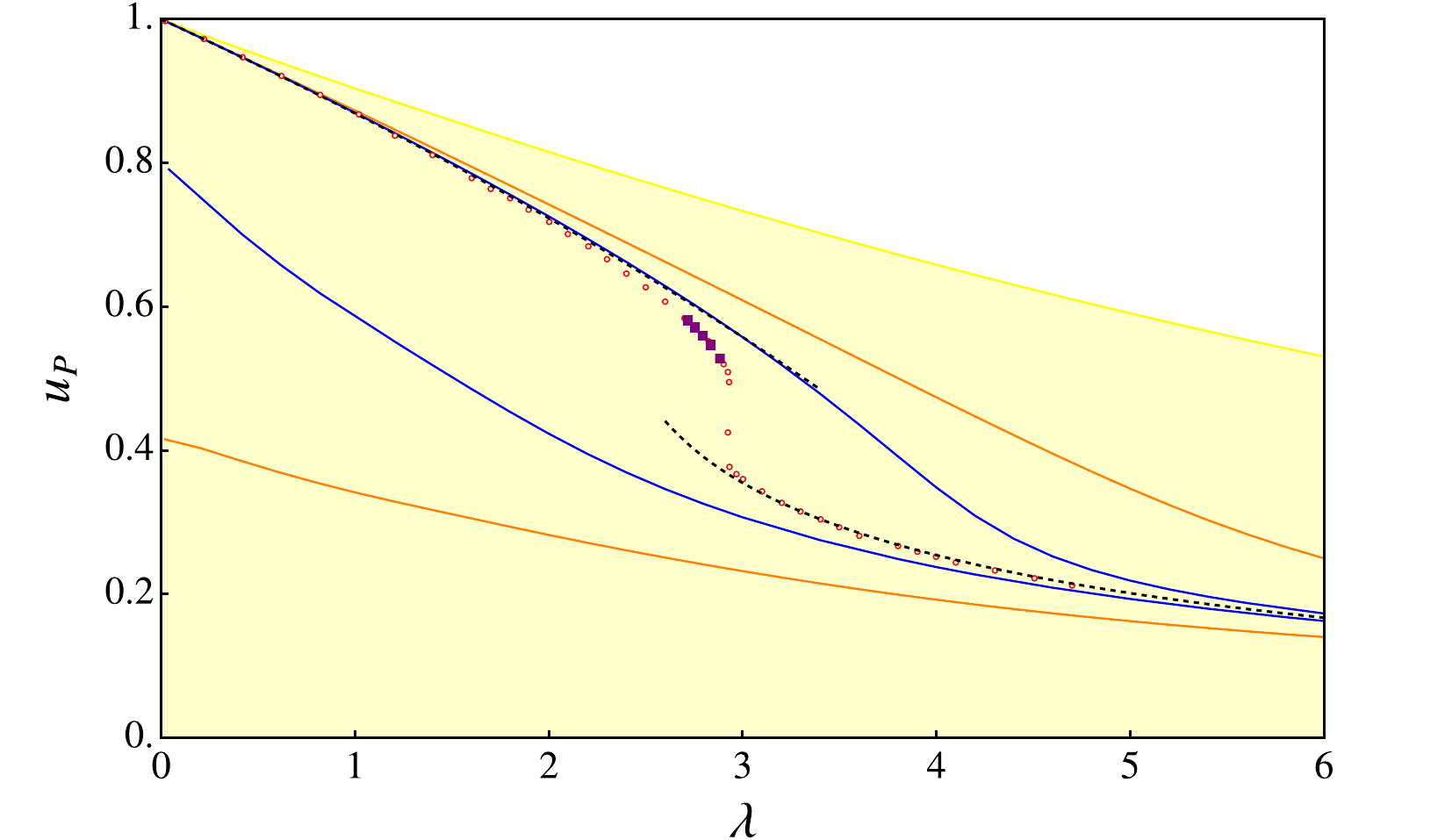}
\caption{Our bootstrap results for upper and lower bounds on plaquette average in \(4D\) LGT \eqref{eq: action}:    for \(L_{\mathrm{max}}=8 \) (yellow domain) for  \(L_{\mathrm{max}}=12\) (orange curves) and  \(L_{\mathrm{max}}=16\) (blue curves). Red circles represent the MC data for \(SU(10)\) LGT (with 5 purple squares for \(SU(12)\)).  Dashed upper and lower lines represent the 3-loop PT~\eqref{PLg}  and strong coupling expansion~\eqref{SCexp}, resp.  }\label{fig:plaquette4D}
\end{figure}   
 
Unlike MC where the result is given up to statistical error bars, the bootstrap methods provides  rigorous inequalities giving  upper and lower bounds on computed physical quantities. These bounds can only improve when increasing the number of bootstrapped variables and constraints on them. 

\begin{figure}
\includegraphics[width=.5\textwidth]{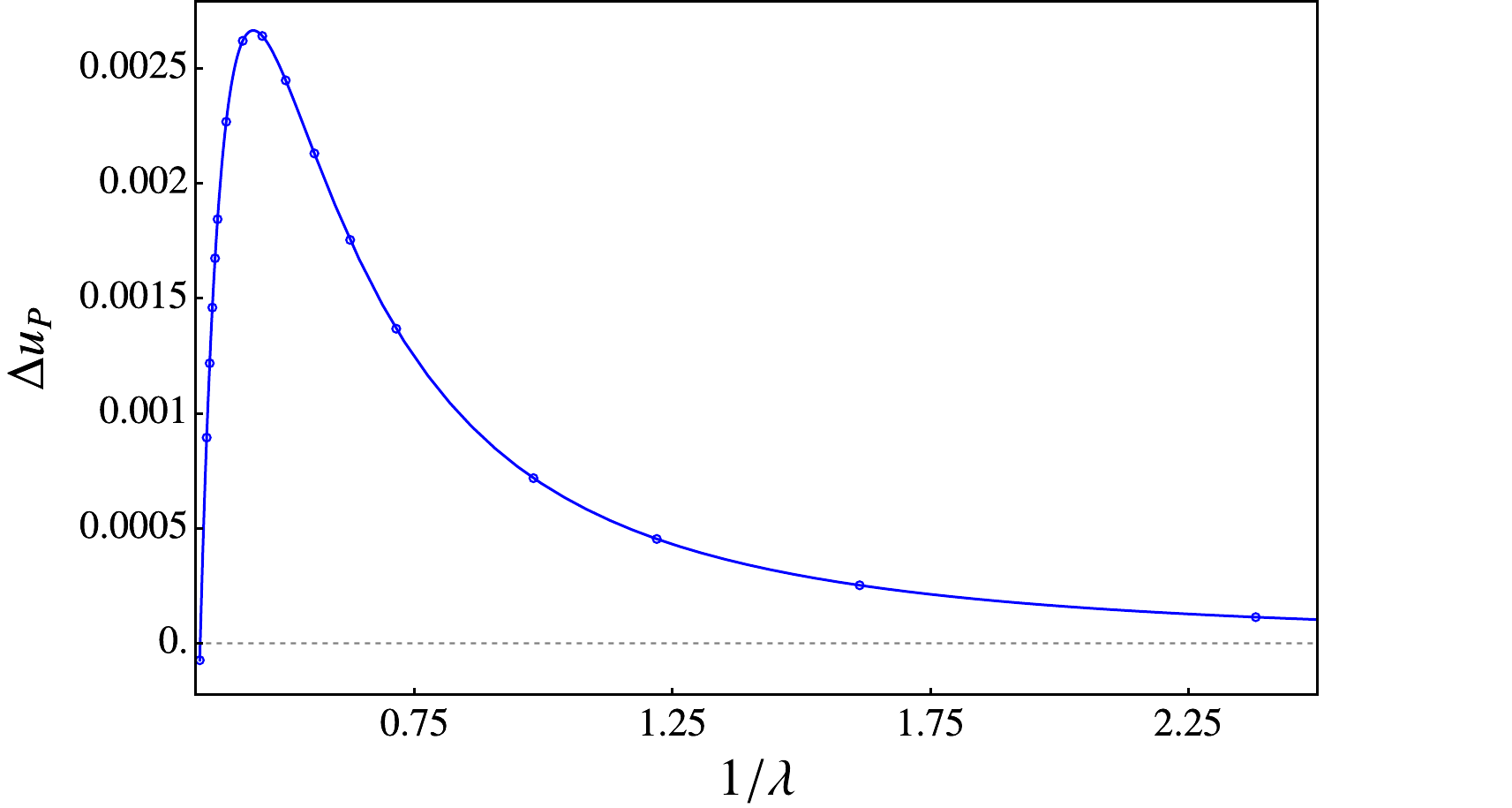}
\caption{The plot \(\Delta u_P\equiv u_P^{\text{boot}}-u_P^{\text{PT}}\) which might capture the non-perturbative values of the gluon condensate  \(\langle \tr \left(F_{\mu\nu}F^{\mu\nu}\right)\rangle\).  }\label{fig:diff}
\end{figure}

Here we develop a powerful numerical bootstrap algorithm for solving the LEs~\cite{Makeenko:1979pb,Migdal:1983qrz} in the lattice YM theory at  \(N_c\to\infty\) and demonstrate it on the computation of plaquette average $u_P=\frac{1}{N_c}\langle\tr U_P\rangle$. Its main ingredients are i)~positivity of correlation matrix of WAs;   ii)~our relaxation procedure of \cite{Kazakov:2021lel}; iii)~positivity of reflection matrices due to lattice symmetries. iv)~symmetry reduction of the positivity conditions. In the supplementary material we worked out a simple example for our general method together with some data points.

Our data (obtained on a single workstation), already for the modest length cutoff \(L_{\mathrm{max}}=16\), look quite encouraging:  as seen on Fig.\ref{fig:plaquette4D}, for \(D=4\), our lower(upper) bounds are quite close to the MC data~\cite{Anderson:2016rcw,Anderson:2018xuq,Athenodorou:2021qvs,Gonzalez-Arroyo:2014dua} in the strong(weak) coupling phase, at least far enough from the phase transition. The upper bound is remarkably close to the 3-loop PT.

On Fig.\ref{fig:diff} we plot the difference of the bootstrap upper bound and  the 3-loop PT:   \(\Delta u_P\equiv u_P^{\text{boot}}-u_P^{\text{PT}}\) as function of \(1/\lambda\). It might capture the non-perturbative effect for the gluon condensate \(\langle \tr \left(F_{\mu\nu}F^{\mu\nu}\right)\rangle\)  -- in principal a measurable observable~\cite{Campostrini:1989uj, Campostrini:1987vz, Bali:2014sja}.  The graph is very smooth and it is positive even slightly beyond the phase transition point \(\lambda_c\simeq 2.9\). ~\footnote{We thank Maxim Chernodub for discussion and suggestions on this subject.}

 \section{Yang-Mills Loop equations at  Large \(N\)  }
 
\begin{figure}
\centering
    \includegraphics[width=.45\textwidth]{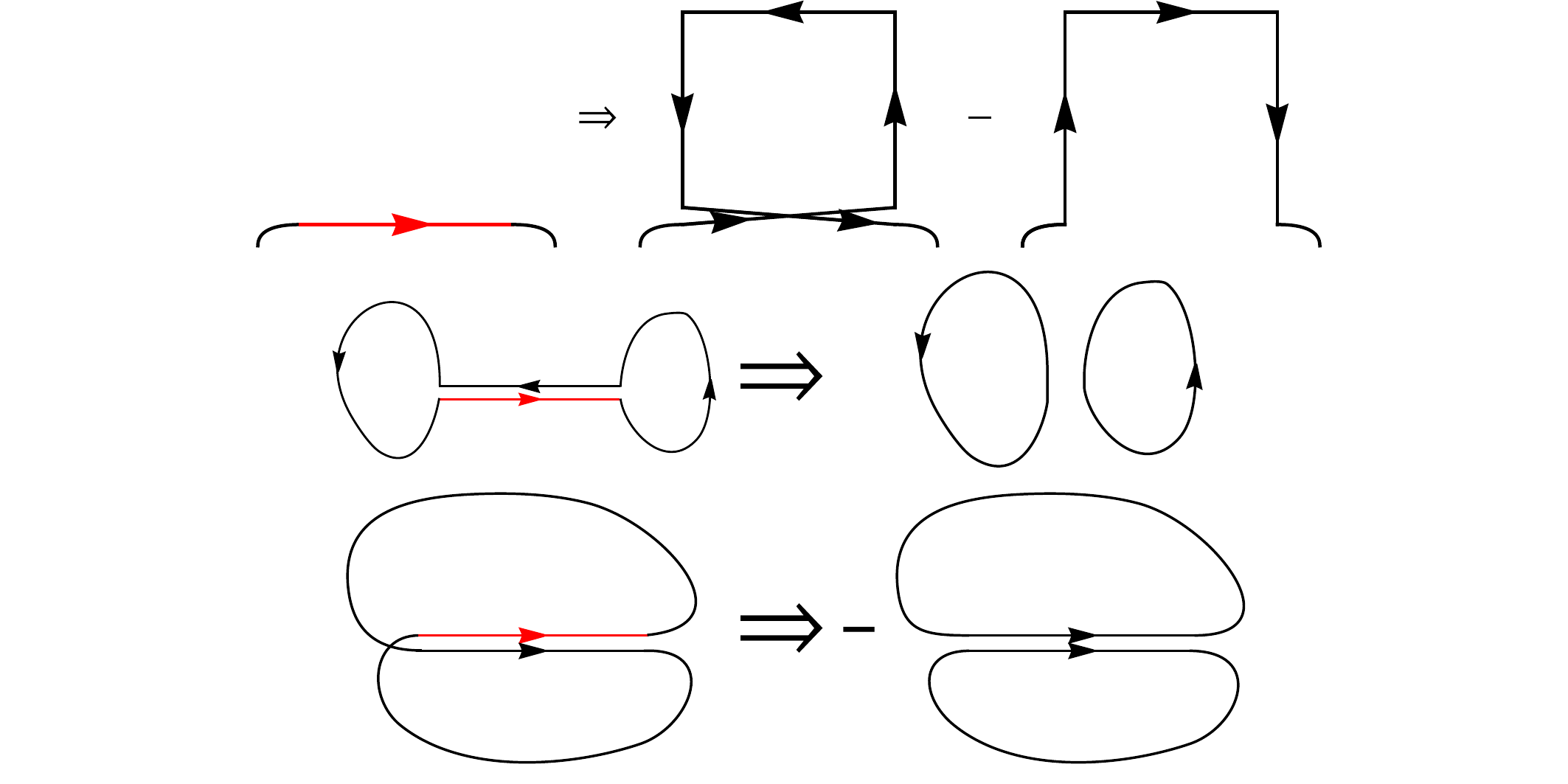}
    \caption{Schematic representation of LEs: The first line shows the variation of a link of Wilson loop in the LHS of Eq.\eqref{MMLE}. The 2nd and 3rd lines show the splitting of the contour along the varied line into two sub-contours, for two different orientations of coinciding links in the RHS of Eq.\eqref{MMLE}.}
 \label{fig:LE}   
\end{figure}
 
 We study the LGT  with the Wilson action~\cite{,Wilson:1974sk}: \begin{equation}\label{eq: action}
 S=-\frac{N_c}{\lambda}\sum_{P} {\rm Re}\,\tr U_P\,
\end{equation}
where \(U_P\) is the product of four unitary link variables around the plaquette \(P\) and we sum over all plaquettes \(P\), including both orientations.  The main quantities of interest in 't~Hooft limit  \(N_c\to\infty\) are the WAs:
\begin{equation}\label{WA}
W[C]=\langle\frac{\tr}{N_c}\prod_{l\in C} U_l\rangle\,.
\end{equation} 
 The matrix product goes over the link variables belonging to the lattice loop \(C\). $W[C]$ are subject to LEs, i.e., the Schwinger-Dyson equations expressing measure invariance  w.r.t.group shifts  \(U_l\to U_l(1+i\epsilon)\). Schematically LE reads: 
\begin{align}\label{MMLE}
&\sum_{\nu\perp\mu}\left(W[C_{l_\mu}\!\!\cdot
\overrightarrow{\delta C^{\nu}_{l_\mu}}]-W[C_{l_\mu}\!\!\cdot\overleftarrow{\delta C^{\nu}_{l_\mu}}]\right)\notag\\ =&\lambda\sum_{\underset{l'\sim l}{l'\in C}}\,\epsilon _{ll'}W[C_{ll'}]\,\,W[C_{{l'l}}]
\end{align}
where the LHS represents the loop operator acting on the link \(l_\mu\) by replacing it with the loop around plaquette \(\overrightarrow{\delta C^{\nu}_{l_\mu}}\) or \(\overleftarrow{\delta C^{\nu}_{l_\mu}}\) (depending on the orientation,  as shown in the first line of Fig~\ref{fig:LE}). The LHS sum goes around all \(2(D-1)\) \(\mu\nu\)-plaquettes orthogonal to the direction \(\mu\). The RHS sum goes over all appearances of the  the same lattice link \(l\) in the loop \(C\). The RHS product corresponds to splitting of the contour \(C\to C_{ll'}\cdot C_{l'l}\), as explained in the 2nd and 3rd lines of Fig.\ref{fig:LE}. Finally, \(\epsilon_{ll'}=\pm 1\) for links \(l\) and \(l'\)  with opposite or colinear orientation, respectively. For more details on LEs, see \cite{Wadia:1980rb, Anderson:2016rcw, Makeenko:2002uj}.

\subsection{Back-track loop equations}
\begin{figure}
    \includegraphics[width=.45\textwidth]{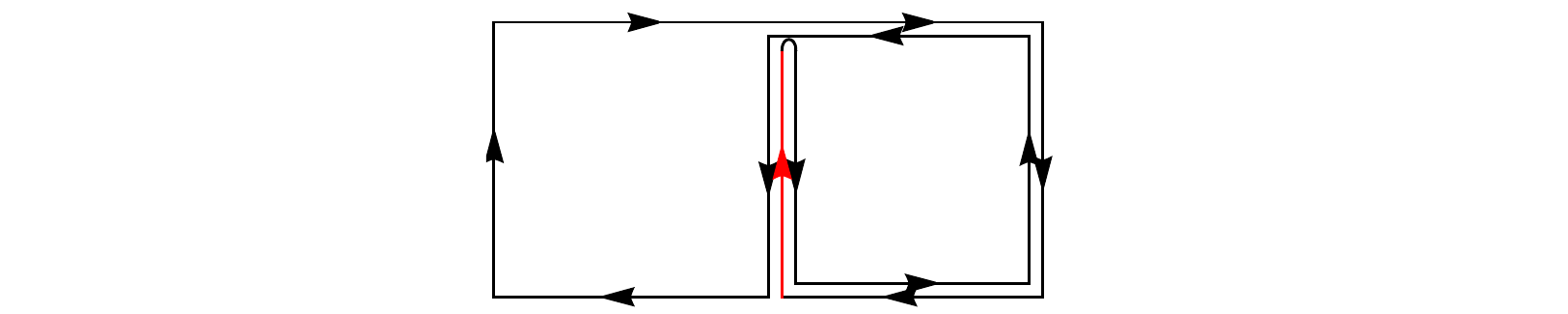}
  \caption{An example of nonlinear ``back-track'' LE.} \label{fig:backtrack}
\end{figure}

To get the full list of the LEs, we also consider the ``back-track'' LEs. They correspond to doing variations on the links at the end of ``back-track" paths originating from the vertices of  Wilson loop. These ``back-tracks''  are equivalent to inserting the identity, but their Schwinger-Dyson variation, can give  independent LEs. Fig~\ref{fig:backtrack} shows an example of ``back-track'' LE:  
\begin{equation}
    \vcenter{\hbox{$\uplong$}}+\vcenter{\hbox{$\backtrackposone$}}-\vcenter{\hbox{$\backtracknegone$}}-\vcenter{\hbox{$\backtracknegtwo$}}=\lambda u_P\vcenter{\hbox{$\uplong$}}.
\end{equation}

The LEs close on single trace WAs \eqref{WA} (due to the large \(N_c\) factorization of color traces),  which means that they behave as classical quantities on the loop space. At \(L_{\mathrm{max}}=16\), we have about \(40,000\) LEs in \(3D\) and about \(100,000\) LEs in  \(4D\), and the back-track LEs constitute more than \(80\%\) of all these LEs.  Around three-quarters of them are linearly independent and only a small minority are non-linear. Finding the solution to these equations is our primary task for any other progress in studying the physical quantities of planar QCD or its \(1/N_c\)  corrections. The problem is very complicated since it is formulated in extremely complex loop space.  We will try to solve it using the bootstrap approach. 


\section{Bootstrap algorithm }

\subsection{Positivity Constraints}
In general, our positivity conditions come from the positivity of possible inner products on the vector space or a subspace of the operators, i.e.
\begin{equation}\label{qform}
    \langle  \mathcal{O} | \mathcal{O}\rangle= \langle  \mathcal{O}^\dagger \mathcal{O}\rangle=  \alpha^{*\mathrm{T}} \mathcal{M}  \alpha\geq 0\Leftrightarrow \mathcal{M}\succeq 0.
\end{equation}
where  $ \mathcal{O}= \sum\alpha_i \mathcal{O}_i$ is an operator with  arbitrary coefficients $\alpha_i$, and $\mathcal{O}_i$ are basis vectors of the operators. 

One of possible adjoint operators $\mathcal{O}^\dagger$ comes from taking the Hermitian conjugation~\cite{Anderson:2016rcw, Lin:2020mme, Kazakov:2021lel}. For a Wilson path, the Hermitian conjugation corresponds to reversing the path. By taking a linear combination of all Wilson paths $0\to x$ (between the points \(0\) and  \(x\)),  with arbitrary coefficients, we can get non-trivial positivity conditions from their inner product. For example, we have only two paths \(0\rightarrow(1,1)\), at \(L_{\mathrm{max}}=2\)
\begin{align} 
     \mathrm{Path}_1=\vcenter{\hbox{$\pathone$}},\quad\mathrm{Path}_2=\vcenter{\hbox{$\pathtwo$}}
\end{align}
 and the positivity condition reads:
\begin{align}
    \kbordermatrix{
    & \mathrm{Path}_1 & \mathrm{Path}_2 & \\
    \mathrm{Path}_1^\dagger & 1 & u_P  \\
     \mathrm{Path}_2^\dagger & u_P & 1  
  }\succeq 0.
\end{align}
This gives \(u_P^2\le 1\), obvious from unitarity. We call the positivity matrices arising from the Hermitian conjugation the correlation matrices.

\begin{figure}
     \centering
     \includegraphics[width=.45\textwidth]{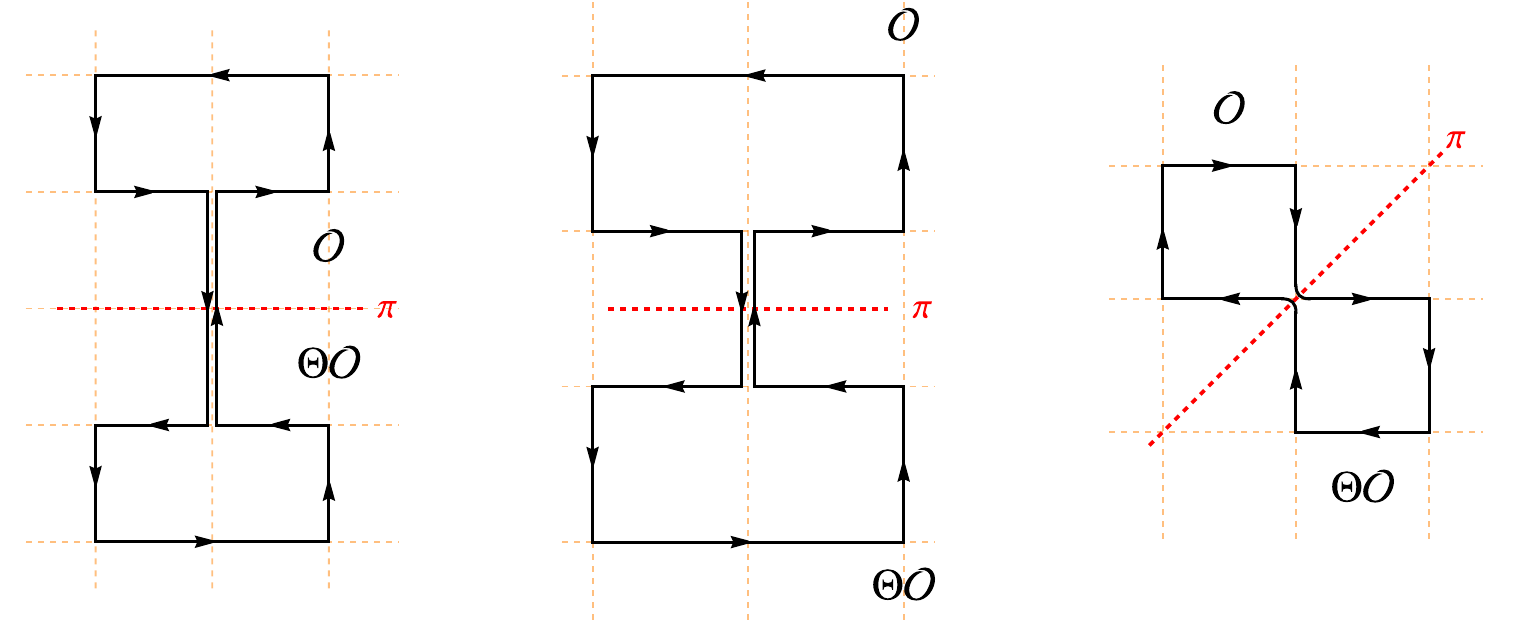}
        \caption{Examples of three reflection symmetries on the lattice allowing new positivity conditions on WAs.}
        \label{fig:ref}
\end{figure}

Apart from Hermitian conjugation, we have  additional reflection positivity conditions where adjoined operators $\mathcal{O}^\dagger$ come from reflection symmetries  ~\cite{Ising}. For LGT, there are three types of reflections w.r.t. different planes: site, link and diagonal reflections~\cite{OSTERWALDER1978440,montvay1997quantum}. Fig~\ref{fig:ref} illustrates the corresponding adjoint paths for these reflections. The importance of 3 new reflection positivity conditions is illustrated on   Fig~\ref{fig:plaquette3D} where we compare \(L_{\mathrm{max}}=12\) bootstrap results with and without reflection positivity. 

For computations in this work,  we consider the full positivity constraint \(0\rightarrow x\) for any possible \(x\) when \(L_{\mathrm{max}}\leq 12\). But for \(L_{\mathrm{max}}= 16\), we consider only the paths \(0\rightarrow 0\) for various positivity matrices since:
\begin{enumerate}
    \item When constructing correlation matrices, all the positivity conditions on the open Wilson paths \(0\rightarrow x\) are already contained in  \(0\rightarrow 0\) correlation matrix for  higher lengths (due to back-trackings).
    \item At \(L_{\mathrm{max}}= 16\), we observe empirically  that the \(0\rightarrow 0\) constraints are computationally the most  efficient. One important reason for that is that the positive matrices corresponding to \(0\rightarrow 0\) are numerically more tractable w.r.t. symmetry reduction that we will discuss below.
\end{enumerate}
\subsection{Convex relaxation}
In general, LEs are non-linear. Here we use the relaxation method to replace all the non-linear LE with linear ones where in the r.h.s. we replace the products of WAs with new variables \(Q_{ij}\),
subject to extra constraints~\cite{Kazakov:2021lel}:
\begin{equation}
    Q_{ij}=W_iW_j\,\,\overset{\text{replace}}{\Longrightarrow}\,\, \begin{pmatrix}
1 & W^{\mathrm{T}}\\
W & Q
\end{pmatrix}\succeq 0.
\end{equation}
Here \(W=\{W_1,W_2,W_3,\dots\}\) denotes the column vector of all inequivalent WAs. Notice that the relaxation matrix has rank=1 precisely when \(Q_{ij}=W_iW_j\).

The LEs combined with the convex relaxation and positivity conditions constitute the constraints of semi-definite programming (SDP). To get  rigorous bounds on WAs, we can maximize or minimize \(u_P\). 

\subsection{Reduction by Symmetry Group}
In conformal  and S-matrix bootstraps, the positivity conditions are well-known to be organized in different spin channels \cite{Rattazzi_2008,Paulos:2017fhb} and different irreducible representations of global symmetry~\cite{Rattazzi:2010yc,He:2018uxa,Cordova:2018uop,Paulos:2018fym}. In parallel to that observation, we can greatly reduce the positivity matrices via the  global lattice symmetries. 

Formally, if we have an invariant group \(G\) preserving the inner product 
\begin{equation}
    \langle (g \circ \mathcal{O}_1)| (g \circ \mathcal{O}_2) \rangle= \langle  \mathcal{O}_1|  \mathcal{O}_2 \rangle ,\, \forall g\in G
\end{equation}
then the positivity condition on the matrix \(\mathcal{M}\) defined in Eq~\ref{qform} can be re-arranged into a block-diagonal form corresponding to the irreps of \(G\).

This well-studied procedure is known  under the name ``Invariant Semidefinite Programming''. Here we refer to a statement~\cite{bachoc2012invariant} directly related to our current problem:
\begin{it}
If the vector space of the paths can be decomposed as a direct sum of irreducible representations \(\mathrm{Rep}_k\) of the invariant group with multiplicity \(m_k\):
\begin{equation}
    V=\bigoplus_{k=1}^D \mathrm{Rep}_k^{\bigoplus m_k},
\end{equation}
then the positivity condition of the inner-product matrix is equivalent to the collection of positivity conditions on the matrices corresponding to each \(\mathrm{Rep}_k\), with matrix dimension \(m_k\times m_k\).
\end{it}

For the correlation matrix  \(0\rightarrow 0\), the invariant group \(G\) is \(B_d \times \mathbb{Z}_2\), where \(B_d\) is the Hyperoctahedral group in \(d\) spacetime dimensions. It acts on a Wilson path by corresponding  rotations and reflections on the spacetime lattice. \(\mathbb{Z}_2\) is the group action reversing the path.

For the reflection positivity matrices  \(0\rightarrow 0\) the invariant groups are subgroups of \(B_d \times \mathbb{Z}_2\), leaving the reflection plane invariant. These invariant subgroups are summarized in Table~\ref{tab:stablizer}.
\begin{table}
\begin{center}
    \begin{tabular}{ |c ||p{2cm}|p{2cm}|p{2cm}|  }
 \hline
 Dimension& Hermitian Conjugation &site\&link reflection &diagonal reflection\\
 \hline
 2& \(B_2\times \mathbb{Z}_2\) & \(\mathbb{Z}_2\times \mathbb{Z}_2\) & \(\mathbb{Z}_2\times \mathbb{Z}_2\)   \\
 3& \(B_3\times \mathbb{Z}_2\) & \(B_2\times \mathbb{Z}_2\) & \(\mathbb{Z}_2^3\)  \\
 4& \(B_4\times \mathbb{Z}_2\) & \(B_3\times \mathbb{Z}_2\) & \(B_2\times \mathbb{Z}_2^2\)\\
 \hline
\end{tabular}
\caption{Invariant groups of   correlation and reflection matrices.}
\label{tab:stablizer}
\end{center}
\end{table}

Implementing this symmetry reduction is similar to projecting the physical state w.r.t spin and parity in conformal or $S$ matrix bootstrap. Practically we do the following steps:
\begin{enumerate}
    \item Find a specific realization of every irrep of the invariant group using GAP software~\cite{GAP4}.
    \item Use the algorithm initiated in \cite{xu2021computation} to find an equivalent real representation (if the irrep by GAP is complex).
    \item  To decompose into such irreps, we use the projector to \(\mathrm{Rep}_k\) ~\cite{serre1977linear}  :
\begin{equation}
    p_{\alpha\alpha, k}=\frac{\dim (\mathrm{Rep}_k)}{\dim G}\sum_{g\in G} r_{\alpha \alpha}(g^{-1}) g
\end{equation}
 Here \(r_{\alpha \beta}\) is a matrix element of a real representation identified at the step 2, and  \(\alpha,\, \beta=1,2,\dots,\dim(\mathrm{Rep}_k)\). Taking \(\alpha=1\), \(P_k=p_{11,k}\) gives us a projector to \(\mathrm{Rep}_k\).
\end{enumerate}

\subsection{Selection of multiplets of Wilson paths}

The Wilson paths form different multiplets of the invariant group. Within each multiplet, the symmetry group permutes different Wilson paths. When constructing the positivity matrices, some multiplets are more important than others. We kept only the most important multiplets. More precisely, several WAs are not related to other WAs by the LEs, such as the \(4\times 4\) square Wilson loop at \(L_{\mathrm{max}}=16\). We believe that such WAs and the open Wilson paths out of which they are constructed are relatively unimportant. 

As an example, take the correlation matrix for the paths \(0 \rightarrow 0\) at \(3D\) and \(L_{\mathrm{max}}=16\). It has a huge size \(6505\times6505\).  After the symmetry reduction and selection of the multiplets, the positivity of the correlation matrix reduces to postivity of 20 smaller matrices, each corresponding to its irrep, with  sizes: 
\begin{equation}
    \begin{split}
        &38, 15, 25, 18, 62, 33, 68, 75, 56, 78,\\
        & 22, 18, 34, 15, 56, 33, 57, 76, 69, 73
    \end{split}
\end{equation}
So the SDP gets greatly simplified.

\begin{figure}[htp]
    \includegraphics[width=.5\textwidth]{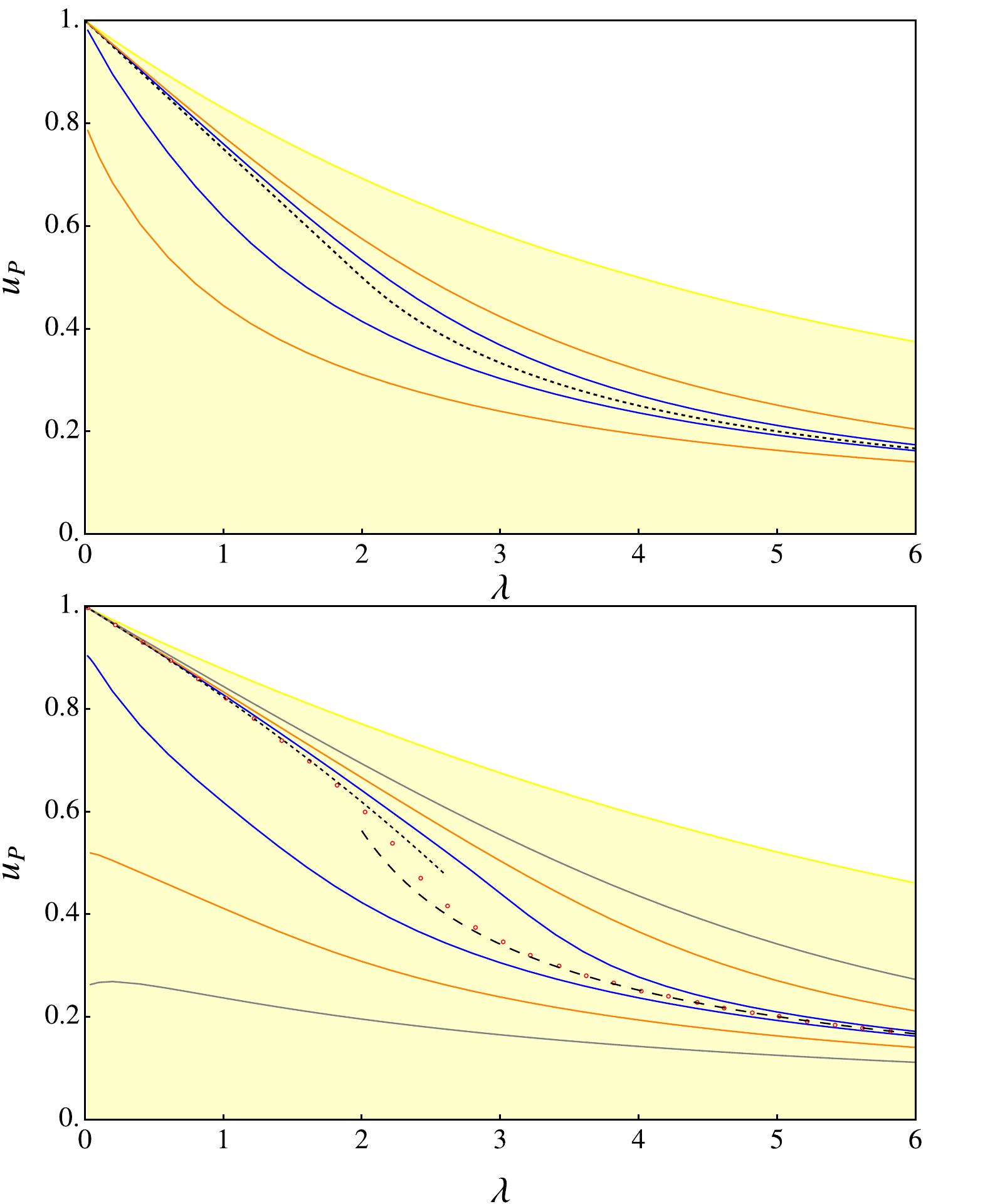}
    \caption{\(u_P\) for \(2D\) (upper) and \(3D\) (lower) LGT: the upper and lower bounds from  our bootstrap at \(L_{\mathrm{max}}=8 \) (yellow region), \(L_{\mathrm{max}}=12\) (orange curves) and  \(L_{\mathrm{max}}=16\) (blue curves). The   \(3D\)  and    \(L_{\mathrm{max}}=12\)  result without reflection positivity (gray curve) is much less constraining. The  line of red circles represents the MC data for \(SU(10)\) LGT. The dashed black curve in \(2D\) plot is the exact solution~\eqref{exact2D}. The dashed black curve in \(3D\) plot is the 3-loop PT result~\cite{Panagopoulos:2006ky}. } \label{fig:plaquette3D}
    
\end{figure}

\section{Discussion of  results}  
 
Here we present the results of computation of plaquette average \(u_P(\lambda)\) for various bare couplings \(\lambda\)  in LGT in \(2D\), \(3D\), and \(4D\). On Fig~\ref{fig:plaquette4D} we compare our bootstrap data in \(4D\) with MC for \(SU(10)\)~\cite{Anderson:2018xuq} and \(SU(12)\)~\cite{Athenodorou:2021qvs} LGT, assuming that, with our accuracy, \(N_c=10,12\) are close enough to \(N_c=\infty\). We also compare it  with the known  3-loop  $N_c=\infty$ PT result
~\cite{Alles:1998is}:
\begin{align}\label{PLg}
\begin{small}
u_P=1-\frac{\lambda }{8}-0.005107 \lambda ^2-0.000794 \lambda ^3+\mathcal{O}(\lambda^4)
\end{small}
\end{align}
as well as with the  SC expansion valid in the SC phase, beyond the 1st order phase transition point \(\lambda_c\simeq2.9\) \cite{Drouffe:1983fv}:
\begin{equation}\label{SCexp}
    \begin{small}
u_P=\frac{1}{\lambda }+\frac{4}{\lambda ^5}+\frac{60}{\lambda ^9}+\frac{136}{\lambda
^{11}}+\frac{1092}{\lambda ^{13}}+\mathcal{O}(\lambda^{-15}).
\end{small}
\end{equation}

The bootstrap bounds on \(u_P\) given for  \(L_{\mathrm{max}}=8,12,16\) on Fig.~\ref{fig:plaquette4D} are quickly improving with the increase of cutoff. The physically most interesting WC phase is much better described by the upper bound. Moreover, we see that the upper bound nicely reproduces the  3-loop PT~\eqref{PLg} for a large range of coupling, even beyond the phase transition point where PT is, strictly speaking, not valid. However,  comparing these results to  MC data, we see that it is not yet so good at capturing  the departure from the PT in the interval  \((2.4, \,2.8)\) where the MC data of~\cite{Athenodorou:2021qvs}  (given by black squares on
 Fig.\ref{fig:plaquette4D}) were used to compute the masses and the string tension. We expect a significant improvement for this range in our data if we reach \(L_{\mathrm{max}}=20 \text{ or even } 24\). However, this will certainly demand much bigger computational resources.

Finally, we briefly discuss \(2D\)  and \(3D\) cases. For \(2D\) LGT the plaquette average  can be computed exactly~\cite{Gross:1980he, Wadia:2012fr} (as well as any loop average, see~\cite{Kazakov:1981sb}):
\begin{equation}\label{exact2D}
u_P=\begin{cases}1-\frac{\lambda}{4},\quad &\text{for}  \lambda\le2 \\
\frac{1}{\lambda},\,\quad\qquad &\text{for}  \lambda\ge 2 \\
\end{cases}
\end{equation}
This example was important for both checking our algorithm and for observing how fast our bootstrap data approach the exact result when increasing \(L_{\mathrm{max}}\). The results are presented on Fig.~\ref{fig:plaquette3D}.  For physically interesting  and   challenging case of \(3D\)  LGT, we compare on Fig.~\ref{fig:plaquette3D}  our bootstrap bounds at \(L_{\mathrm{max}}=8,12,16 \)  with the MC data~\cite{Anderson:2018xuq} as well as with the known 3-loop PT~\cite{Panagopoulos:2006ky}  and SC~\cite{Drouffe:1983fv} results. We observe a reasonably fast approach of bootstrap bounds to the MC data when increasing \(L_{\mathrm{max}}\), but they are not as close to PT  as in \(4D\) case.

We employ in LEs the WAs up to the maximal length \(L_{\mathrm{max}}\), so it can be considered as our IR cutoff.   The physical scale \(l_{ph}\) (set by inverse mass or square root of string tension) should, ideally, satisfy    \(1\ll\frac{ l_{ph}}{a_L}\ll L_{\mathrm{max}}\), where \(a_L\) is the lattice spacing. In this paper, we have, for the best of our data, \(L_{\mathrm{max}}=16\) which suggests that the window for the scale of measurable physical quantities in lattice units should be roughly \(2\lesssim \frac{l_{ph}}{a_L}\lesssim 6\) (compare it to Table~8 of~\cite{Athenodorou:2021qvs} where the IR cutoff is set by the size of space-time torus, typically in the range \(10\text{ to }16\), and the typical physical length is set by string tension \( 3\lesssim\frac{l_{ph}}{a_L}\lesssim6\)). 

We conclude that, even though the currently achieved values of \(L_{\mathrm{max}}\) in our bootstrap approach may be not sufficient to match the precision and scope of the MC experiments,  especially for the confinement sensitive physics (glueball masses, string tension, etc.), our results  give good hopes on a considerable improvement when augmenting \(L_{\mathrm{max}}\). Moreover, for sufficiently small couplings our upper bound data are already at least as good as MC.

\section{Prospects}

The bootstrap procedure proposed here has a clear perspective for improving our results and advancing towards the computation of interesting observables.  In particular, by choosing objective functions other than $u_P$, one can hope to get a better estimate for all involved physical quantities.  
For our current implementation at  \(L_{\mathrm{max}}=16\), every data point takes \(\backsim 20\) hours of CPU time for \(4D\), and only half an hour for   \(3D\)~\footnote{All the SDPs are solved by MOSEK.\cite{mosek}}. First, we want to increase the cutoff to \(L_{\mathrm{max}}=20\) and even to \(L_{\mathrm{max}}=24\). This will certainly need supercomputer power. From our current results, we expect a quick narrowing of our bounds to the accuracy comparable to, or even better than MC (without its toll of statistical and systematic errors). Furthermore, since in the 't~Hooft limit we don't have internal fermion loops, we can try to find the quark condensate and hadron masses by simply summing up the WAs with the spinorial factors for the relevant one- and two-point functions. The $1/N_c$ corrections (which might be small enough even for the physical $N_c=3$ case) seem to be not insurmountable tasks since they are subject to {\it linear}  LEs~\cite{Migdal:1983qrz}, with coefficients given by the solution of LE~\eqref{MMLE}. One of such problems is the computation of glueball masses from the connected correlator of two small Wilson loops. One can also try to bootstrap directly the $N_c=3$ YM theory, where the absence of large factorization could be compensated by multiple functional relations between WAs, absent for $N_c=\infty$ case.

%
%

\begin{acknowledgments}
\section*{Acknowledgments}
We thank Benjamin Basso, Maxim Chernodub, Alexander Gorsky, David Gross, Shota Komatsu, Marina Krstic-Marinkovic, Martin Kruczenski, Robert Pisarski, Jiaxin Qiao and Junchen Rong for useful discussions. V.K. thanks KITP of California University at Santa Barbara for kind hospitality during the work on this project. 
\label{sec:acknowledgments}
\end{acknowledgments}

\bibliographystyle{apsrev4-2}
\bibliography{YMbootstrap}

\newpage

\onecolumngrid

\appendix

\setcounter{equation}{0}

\renewcommand{\theequation}{S\arabic{equation}}
\renewcommand\thefigure{S\arabic{figure}}

{\begin{center} {\bf \large Supplementary Material for ``Bootstrap for Lattice Yang-Mills theory"} \\
\vspace{0.3cm}
Vladimir Kazakov, Zechuan Zheng\end{center}}

\bigskip

\section{A worked-out example}

This section gives a step-by-step solution of $L_{\mathrm{max}}=8$ bootstrap bound for 2-dimensional lattice Yang-Mills theory in the large $N_c$ limit. For simplicity, here, we will only take into account positivity conditions from Wilson lines starting from the origin and ending at the origin. In this situation, up to rotations and reversing the line, we have only one multiplet under symmetries:
\begin{figure}[h]
  \includegraphics[width=.1\textwidth]{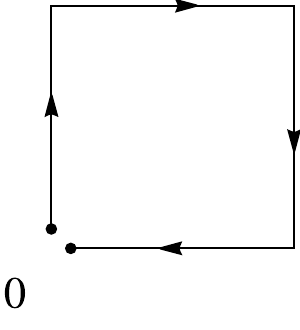}
\label{fig:line}
\end{figure}   

Taking the inner product between the different Wilson lines in this multiplet, we can construct the positivity matrices corresponding to Hermitian conjugation (correlation matrices) and space reflection (reflection matrices). For the correlation matrix, we have:

\begin{equation}\label{pos}
  \begin{pNiceMatrix}[first-row,first-col]
    & \vcenter{\hbox{$\mathbb{I}$}} & \lineone& \linetwo& \linethree& \linefour& \linefive& \linesix& \lineseven& \lineeight & \\
   \vcenter{\hbox{$\mathbb{I}\quad$}} & 1 & \wone & \wone & \wone & \wone & \wone & \wone & \wone & \wone \\
   \linefive & \wone & 1 & \wtwo & \wtwo & \wsix & \wthree & \wfour & \wfour & \wfive \\
   \lineseven& \wone & \wtwo & 1 & \wsix & \wtwo & \wfour & \wfive & \wthree & \wfour \\
   \linesix & \wone & \wtwo & \wsix & 1 & \wtwo & \wfour & \wthree & \wfive & \wfour \\
   \lineeight & \wone & \wsix & \wtwo & \wtwo & 1 & \wfive & \wfour & \wfour & \wthree \\
   \lineone &\wone & \wthree & \wfour & \wfour & \wfive & 1 & \wtwo & \wtwo & \wsix \\
   \linethree & \wone & \wfour & \wfive & \wthree & \wfour & \wtwo & 1 & \wsix & \wtwo \\
   \linetwo & \wone & \wfour & \wthree & \wfive & \wfour & \wtwo & \wsix & 1 & \wtwo \\
   \linefour & \wone & \wfive & \wfour & \wfour & \wthree & \wsix & \wtwo & \wtwo & 1 \\ 
 \end{pNiceMatrix}\succeq 0.
\end{equation}
The row above the matrix is the Wilson lines for the inner products, and the column left to the matrix is their Hermitian conjugation. The $\mathbb{I}$ denotes the length 0 trivial Wilson line. The inner product is defined by joining the path and its Hermitian conjugation. This matrix contains six independent Wilson loops:
\begin{equation}
  \scalebox{2}{$\wone\quad\wtwo\quad\wthree\quad\wfour\quad\wsix\quad\wfive\quad$}
\end{equation}
We denote them as:
\begin{equation}
  \mathcal{W}_1,\quad\mathcal{W}_2,\quad\mathcal{W}_3,\quad\mathcal{W}_4,\quad\mathcal{W}_5,\quad\mathcal{W}_6\quad
\end{equation}
respectively. (Here, the $\mathcal{W}_1$ is the $u_P$ in the main text.)

We also note to the reader that although at length 8, there are Wilson loops like the $2\times 2$ Wilson loop doesn't appear in the correlation matrix. We will also see that this is true for the loop equations and the reflection matrices, i.e., this will be an SDP with six variables. 

There are two reflection matrices corresponding to site reflection and diagonal reflection. In terms of $\mathcal{W}$ variables, they read:
\begin{equation}
  \left(
\begin{array}{ccccc}
 1 & \mathcal{W}_1 & \mathcal{W}_1 & \mathcal{W}_1 & \mathcal{W}_1 \\
 \mathcal{W}_1 & \mathcal{W}_2 & \mathcal{W}_6 & \mathcal{W}_4 & \mathcal{W}_5 \\
 \mathcal{W}_1 & \mathcal{W}_6 & \mathcal{W}_2 & \mathcal{W}_5 & \mathcal{W}_4 \\
 \mathcal{W}_1 & \mathcal{W}_4 & \mathcal{W}_5 & \mathcal{W}_2 & \mathcal{W}_6 \\
 \mathcal{W}_1 & \mathcal{W}_5 & \mathcal{W}_4 & \mathcal{W}_6 & \mathcal{W}_2 \\
\end{array}
\right) \,\mathrm{and}\,\left(
  \begin{array}{ccc}
   1 & \mathcal{W}_1 & \mathcal{W}_1 \\
   \mathcal{W}_1 & \mathcal{W}_5 & \mathcal{W}_6 \\
   \mathcal{W}_1 & \mathcal{W}_6 & \mathcal{W}_5 \\
  \end{array}
  \right),
\end{equation}
The reflection matrix corresponds to link reflection is trivial here. Although not evident, at the present bootstrap cutoff, the positivity of reflection matrices won't improve the constraint. So we won't further discuss these matrices (including their symmetry reduction) and do the optimization only under the positivity of the correlation matrix.
\begin{figure}[h]%
  \centering
  \qquad\qquad
  \includegraphics[width=2.5cm]{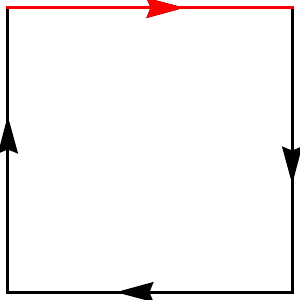} 
  \qquad
  \includegraphics[width=5cm]{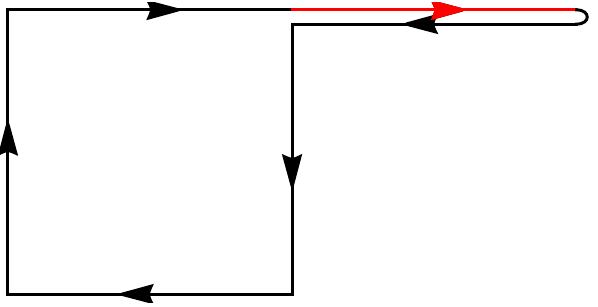} 
  \caption{Loop equations for $\Lambda=4$.}%
  \label{fig:loop}%
\end{figure}

There are two loop equations, as shown in Fig~\ref{fig:loop}, the right one is a back-track loop equation. In terms of $\mathcal{W}$ variables, they read:
\begin{align}
    -\mathcal{W}_2+\mathcal{W}_3+\mathcal{W}_4-1&=-\lambda\mathcal{W}_1 \nonumber\\
  \mathcal{W}_2-\mathcal{W}_4-\mathcal{W}_5+\mathcal{W}_6&=0\label{loop}
\end{align}
Finally, we do the symmetry reduction on the correlation matrix. This process will make the matrix factorize to a series of smaller matrices corresponding to each irreducible representation of the global symmetry group. We stress that this step is purely group theoretical, it doesn't change the positivity constraint itself, but it makes our algorithm orders of magnitude faster.

In our current dimension, as shown in the main text, the global symmetry is actually $B_2\times \mathbb{Z}_2$. The $B_2$ group is the cubic discrete version of $O(2)$, and isomorphic to the Dihedral group $D_4$. The $\mathbb{Z}_2$ group corresponds to the symmetry of reversing the Wilson loop. The $\mathbb{Z}_2$ group is the Charge Conjugation symmetry on the lattice system, ensuring that all the Wilson loop averages are real numbers. 

The Wilson lines above the matrix Eq~\ref{pos} actually form a multiplet for the global symmetry group. We can decompose this multiplet into each irreducible component. And from the discussion in the main text, the Wilson lines in different irrep are orthogonal in the corresponding inner product.

For the group $B_2\times \mathbb{Z}_2$, there are ten irreducible representations:
  \begin{align}
    (A_1, +1) && (A_2, +1) &&(B_1, +1)&&(B_2, +1)&& (E, +1)\\
    (A_1, -1) && (A_2, -1) &&(B_1, -1)&&(B_2, -1)&& (E, -1)
  \end{align}
The first index is the standard notation for irreducible representation of Dihedral group $ D_4\backsimeq B_2$, and the second one is the parity of Charge Conjugation. Using the algorithm stated in the main text, we can project the multiplet to different irreps:
\begin{equation}
  \begin{split}
    (A_1, +1):\, & \mathbb{I}, \quad \frac{1}{8}( \lineone + \linetwo + \linethree + \linefour + \linefive + \linesix + \lineseven + \lineeight ) \\
    (B_2, +1):\, & \frac{1}{8} (-\linethree-\lineseven-\linesix-\linetwo+\lineone+\linefive+\lineeight+\linefour)\\
    (E, +1):\, & \frac{1}{4} (-\linethree-\lineseven+\linesix+\linetwo)\\
    (B_1, -1):\,  & \frac{1}{8} (-\lineone-\linethree-\linetwo-\linefour+\linefive+\lineseven+\linesix+\lineeight)\\
    (A_2, -1):\,  & \frac{1}{8}(-\lineone-\lineseven-\linesix-\linefour+\linethree+\linefive+\lineeight+\linetwo)\\
    (E, -1):\,  & \frac{1}{4} (-\lineone-\linefive+\lineeight+\linefour)
  \end{split}
\end{equation}
We note that for the two irreps correspond to two $E$, there are other two vectors, but they will give the same positivity matrices. They are fully redundant. Taking the inner product between the vectors in the same irrep, we get the following positivity condition:
\begin{align}
  \left(
\begin{array}{cc}
 1 & \mathcal{W}_1 \\
 \mathcal{W}_1 & \frac{1}{4} \mathcal{W}_2+\frac{1}{8} \mathcal{W}_3+\frac{1}{4} \mathcal{W}_4+\frac{1}{8} \mathcal{W}_5+\frac{1}{8} \mathcal{W}_6+\frac{1}{8} \\
\end{array}
\right)\succeq 0,\nonumber\\
 -\frac{1}{4} \mathcal{W}_2+\frac{1}{8} \mathcal{W}_3-\frac{1}{4} \mathcal{W}_4+\frac{1}{8} \mathcal{W}_5+\frac{1}{8} \mathcal{W}_6+\frac{1}{8} 
\geq 0,\nonumber\\
 -\frac{1}{4} \mathcal{W}_3+\frac{1}{4} \mathcal{W}_5-\frac{1}{4} \mathcal{W}_6+\frac{1}{4} \geq 0,\nonumber\\
 \frac{1}{4} \mathcal{W}_2-\frac{1}{8} \mathcal{W}_3-\frac{1}{4} \mathcal{W}_4-\frac{1}{8} \mathcal{W}_5+\frac{1}{8} \mathcal{W}_6+\frac{1}{8} \geq 0,\nonumber\\
 -\frac{1}{4} \mathcal{W}_2-\frac{1}{8} \mathcal{W}_3+\frac{1}{4} \mathcal{W}_4-\frac{1}{8} \mathcal{W}_5+\frac{1}{8} \mathcal{W}_6+\frac{1}{8} \geq 0, \nonumber\\
 \frac{1}{4} \mathcal{W}_3-\frac{1}{4} \mathcal{W}_5-\frac{1}{4} \mathcal{W}_6+\frac{1}{4}\geq 0.\label{sym}
\end{align}
The reader could verify that these new positivity conditions are equivalent to the Eq~\ref{pos}. Still, as SDP conditions, they are much more (several orders of magnitude for higher order bootstrap problems) efficient.

By the loop equation Eq~\ref{loop} and the factorized positivities Eq~\ref{sym}, we can maximize or minimize $\mathcal{W}_1$ to get the upper and lower bounds. For $\lambda=2$, we have:
\begin{equation}
  0\leq \mathcal{W}_1\leq 0.69300
\end{equation}

We stress that all the loop equations are linear at this lowest order of bootstrap. We must implement the convex relaxation described in the main text to deal with the quadratic terms for higher order bootstrap.

\section{Selected four-dimensional data}
In this section we collect some selected data from the $4D$ bootstrap plot. Table~\ref{tab: num} compares the bootstrap upper bound on \(u_P\) with the perturbation and MC data in the weak-coupling phase at several specific couplings. 

\begin{table}[h]
  \begin{center}
      \begin{tabular}{ |c ||p{2cm}|p{2cm}|p{2cm}|  }
   \hline
   \(\lambda\)& \(u
  _{p,\mathrm{max}}\) &Perturbation &Monte Carlo\\
   \hline
   0.01  & 0.99750    &0.997498&   0.9975\\
   0.31 &   0.92060  & 0.920348   &0.9212\\
   0.51 &0.86706  & 0.866344&  0.8676\\
   1.35913 & 0.60837& 0.606530&  0.58043063\\
   1.37733 & 0.60199  &0.600313& 0.57087665 \\
   1.39765 & 0.59480  &0.593336& 0.55936304 \\
   1.41802 & 0.58752  &0.586302& 0.5464461 \\
   1.44202 & 0.57886  & 0.577965   &0.5275951\\
   \hline
  \end{tabular}
       \caption{ Plaquette average at various 't~Hooft couplings \(\lambda\): bootstrap, perturbation theory and MC results.  }
      \label{tab: num}
  \end{center}
  \end{table}

\end{document}